\documentclass[prl,twocolumn,nofootinbib,superscriptaddress,showkeys,citeautoscrip]{revtex4}
\pdfoutput=1
\usepackage[usenames,dvipsnames]{color}
\usepackage{graphicx}
\usepackage{amsmath}
\usepackage{amsfonts}
\usepackage{amssymb}
\usepackage{dsfont}
\usepackage{dcolumn}
\usepackage{bm}
\usepackage{slashed}
\usepackage{pstricks}
\usepackage{slashed}
\usepackage{multirow}
\usepackage{array}
\usepackage{rotating}
\usepackage{xcolor}
\usepackage{epstopdf}
\usepackage{fancybox}
\usepackage{ulem,fancyvrb}

\newcolumntype{x}[1]{
{\centering}p{#1}}%
\newcommand{\tnhl}{\tabularnewline\hline}

\newcommand{\GeV}      {~\mathrm{GeV}}
\newcommand{\pb}      {~\mathrm{pb}}

\newcommand{\beqn}{\begin{eqnarray}}
\newcommand{\eeqn}{\end{eqnarray}}
\newcommand{\be}{\begin{equation}}
\newcommand{\ee}{\end{equation}}
\newcommand{\non}{\nonumber \\}

\newcommand{\mathsym}[1]{{}}

\def \n34{\tilde{\chi}^{0}_{3,4}}

\def\met100{\slashed{E}_T\geq 100 \GeV}

\newcommand{\gappeq}{\mathrel{\rlap {\raise.5ex\hbox{$>$}}
{\lower.5ex\hbox{$\sim$}}}}
\newcommand{\lappeq}{\mathrel{\rlap{\raise.5ex\hbox{$<$}}
{\lower.5ex\hbox{$\sim$}}}}

\def\zp{Z^{\prime}}

\def\met{\slashed{E}_{T}}

\begin{document}
\title{An Explanation of the CDF Dijet Anomaly 
within a $U(1)_X$ Stueckelberg Extension }

\author{Zuowei~Liu}
\affiliation{C.N.\ Yang Institute for Theoretical Physics, 
Stony Brook University, Stony Brook, NY 11794, USA}

\author{Pran~Nath}
\affiliation{Department of Physics, Northeastern University,
 Boston, MA 02115, USA}

\author{Gregory~Peim}
\affiliation{Department of Physics, Northeastern University,
 Boston, MA 02115, USA}


\begin{abstract}
We discuss the  recent excess seen by the CDF Collaboration in the dijet invariant mass distribution
produced in association with a  $W$~boson. We analyze the possibility of such a signal within the context
of a $U(1)_X$ Stueckelberg extension of the Standard Model where the new gauge boson couples only 
to quarks. In addition to the analysis of the $Wjj$ anomaly we also discuss the production of $Zjj$ and $\gamma jj$  at the Tevatron.  The analysis is then extended to the Large Hadron Collider with $\sqrt{s}=7~{\rm TeV}$ and predictions for the dijet signals are made. 
\end{abstract}

\maketitle

Recently, the CDF Collaboration~\cite{CDF} has reported an excess of events in the
invariant mass distribution of jet pairs produced in association with a $W$ 
boson in $p \bar{p}$ collisions at $\sqrt{s}= 1.96$~TeV. 
In this note we analyze this anomaly in the framework of a  Stueckelberg 
$U(1)_X$ extension of the Standard Model~\cite{kn,fln,Feldman:2010wy}.  
The mechanism to explain the anomaly that we propose is different from those discussed 
in the literature~\cite{wjjrefs,ua2pheno,Cheung:2011zt,Wjjzaref,nelson,perezetal}. 
Further, we consider the associated $Zjj$ and $\gamma jj$ production, which is addressed only in~\cite{Cheung:2011zt,Wjjzaref}. 
Additionally, for this framework we study the production of $Wjj$, $Zjj$ and $\gamma jj$ at the Large Hadron Collider at $\sqrt{s}=7~{\rm TeV}$~(LHC7).\\

We begin by extending the Standard Model by the following additional piece in the Lagrangian
\beqn
{\cal L}_1 = &-&\frac{1}{4} X_{\mu\nu}X^{\mu\nu}
+ g_X X_\mu J^\mu_X  \nonumber\\
 &-& \frac{1}{2} (\partial_{\mu}\sigma  
 + M_1 X_{\mu} + M_2 B_{\mu})^2\ .
\label{1}
\eeqn
where  $B_{\mu}$ $\left(X_{\mu}\right)$ is the gauge boson associated with the gauge group $U(1)_Y$ $\left(U(1)_X\right)$, where $Y$ refers to the hypercharge. 
The Lagrangian of Eq.(\ref{1}) is invariant under hypercharge $U(1)_Y$ transformations 
$\delta_Y B_{\mu} = \partial_{\mu}\lambda_Y$, $\delta_Y X_{\mu} =0$ and
$\delta_Y \sigma = - M_2 \lambda_Y$, and under the $U(1)_X$ transformations 
$\delta_X X_{\mu} = \partial_{\mu}\lambda_X$, $\delta_X B_{\mu} =0$ and 
$\delta_X \sigma = - M_1 \lambda_X$.
Thus   there are three neutral gauge bosons  in the extended Lagrangian, 
${\cal L}={\cal L}_{SM} + {\cal L}_1$, which
 are $X_{\mu}$, $B_{\mu}$ and  $A_{\mu}^3$, where  $A_{\mu}^3$
  is the third component of the $SU(2)_L$ gauge multiplet $A_{\mu}^a$~($a=1,2,3$).\\

\begin{figure}[h!]
\includegraphics[scale=0.2]{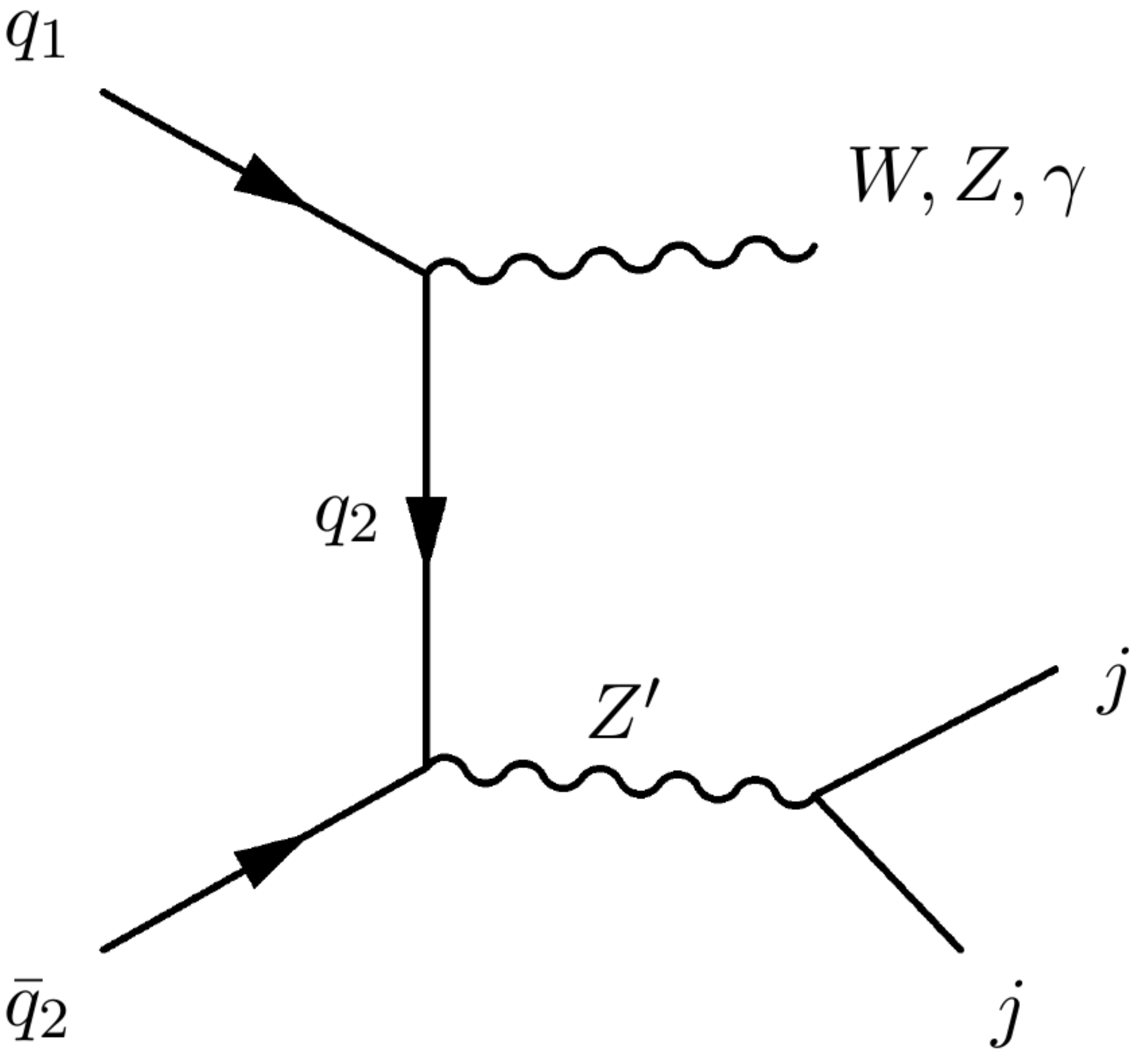}
\caption{\label{feyndiag} Display of the production modes studied in this work
for the $Wjj$, $Zjj$, $\gamma jj$ events.  The direct channel production such as
$q_1\bar q_2 \to W\to W \zp$ is suppressed  as discussed in the text.}
\end{figure}

 We will focus on the neutral current interaction which arises from the couplings of 
 $A_{\mu}^3$, $B_{\mu}$ and $X_{\mu}$,
\beqn
{\cal L}_{\rm NC} = g_2 A_\mu^3 J^{3\mu}_2 + g_Y B_\mu J^\mu_Y + g_X X_\mu J^\mu_X \ ,
\eeqn
where $J^{3\mu}_2$ is the third component of $SU(2)_L$ current, $J^{\mu}_Y$ is the hypercharge
current and $J^{\mu}_X$ is a vector current to which the $U(1)_X$ gauge field $X_{\mu}$ 
couples. 
After spontaneous breaking of the electroweak symmetry one will have, along with Eq.(\ref{1}),
a $3\times 3$ mass matrix which mixes the three neutral gauge fields $A_{\mu}^3, B_{\mu}, X_{\mu}$.
The diagonalization of this mass matrix leads to  a massless neutral state~(the photon), and two 
massive neutral bosons~(the $Z$~boson and the new $\zp$~boson). 
Transforming to the  mass diagonal basis, the couplings of the $Z$ and $\zp $ arising from 
$J^{3\mu}_2$ and $J^{\mu}_Y$ 
are given  by the following
\beqn
{\cal L}_{\rm 2} = &\frac{g_2}{\cos(\theta)}& [
{\rm Z}_\mu (
 \cos(\psi) ( \sin^2(\theta) Q J^\mu_{\rm em} -  J_2^{3\mu} ) \non
 &-&\tan(\phi)\sin(\psi)\sin(\theta)
 ( Q J^\mu_{\rm em} - J_2^{3\mu} ) )
\non
&+&{\rm Z}_\mu'  (
\sin(\psi) ( \sin^2(\theta) Q J^\mu_{\rm em} - J_2^{3\mu} ) \non
&+&\tan(\phi)\cos(\psi)\sin(\theta)
 ( Q J^\mu_{\rm em} - J_2^{3\mu} ) )]\ . 
\label{z1}
\eeqn
Additionally one has the following set of couplings for $Z$ and $\zp $ from $J_X^{\mu}$
\beqn
{\cal L}_{2}'= &g_X& ( \cos\psi \cos\phi -\sin\theta\sin\phi\sin\psi) \zp _{\mu} J^{\mu}_X\nonumber\\
+ &g_X&(-\sin\psi \cos\phi -\sin\theta\sin\phi\cos\psi)Z_{\mu}J^{\mu}_X\ .
\label{z2}
\eeqn
In the above the angles $\phi$ and $\psi$ are given by 
\beqn 
\tan (\phi) &=& \frac{M_2}{M_1},\non
\tan (\psi)& =& \frac{\tan(\theta)\tan(\phi)M_{{\rm W}}^2}
                     {\cos(\theta)(M_{{\rm Z}'}^2-M_{\rm W}^2(1+\tan^2(\theta)))}\ ,
          \eeqn           
               where       
 $\tan(\theta)=\tan(\theta_W)\cos(\phi)$.
 In addition to the above there is also a triple gauge boson vertex with the $\zp WW$ couplings given by
\beqn
\mathcal{L}_{\zp WW}=i g_2 R_{31}[W_{\mu \nu
}^+W^{-\mu}Z^{\prime\nu}+W_{\mu \nu }^-W^{+\nu }Z^{\prime\mu} \non
+W^{+\mu
}W^{-\nu}\zp_{\mu \nu}]\ .\label{wwzp}
\eeqn
where 
\beqn
R_{31}= -\cos(\theta) \sin\psi\ .\label{r31}
\eeqn
We next consider a specific model for $J_{X}^{\mu}$ so that 
$J_X^{\mu} = \sum_q \bar q \gamma^{\mu} q$.
 Now from the electroweak data 
the ratio $M_2/M_1$ is known to be typically small, i.e., $M_2/M_1\ll 1$. For small $M_2/M_1$, 
both $\tan\phi$ and $\tan\psi$ are small, i.e., $\tan\phi, \tan\psi \ll 1$. 
Thus the couplings of the $\zp $ to fermions given by Eq.(\ref{z1}) would be typically much smaller 
compared to the couplings of $\zp $ given by Eq.(\ref{z2}). Further, for the same reason 
$R_{31} \ll 1$, and thus the $\zp WW$ vertex of Eq.(\ref{wwzp}) is significantly suppressed. The implication of the 
above is the following: for the amplitude  $q_1\bar q_2\to W \zp $, the s-channel pole 
contribution via $q_1\bar q_2 \to W\to W\zp $  will be  suppressed  compared to the 
t-channel production of $W\zp $. Thus we focus on the $W\zp $ production via the t-channel exchange, which is illustrated in Fig.~\ref{feyndiag}.
Since by assumption $X_{\mu}$ couples  only to quarks, and since $\zp $ is dominantly $X_{\mu}$
in the limit $M_2/M_1 \ll 1$,  
the  decays of $\zp $ are dominantly to quarks and the leptonic final states, $W\ell^+\ell^-$, are suppressed. Thus in this case   
$W\zp $ production will result in $Wjj$, i.e. a $W$ boson plus dijets. \\

The signal of a baryonic vector boson is constrained by the dijet search at 
colliders. The current Tevatron data constrains the $\zp $ boson with Standard 
Model couplings within the mass range $\in (320,740)\GeV$~\cite{Aaltonen:2008dn}. 
Below $200\GeV$, however, the UA2 experiment~\cite{ua2}
gives a better constraint than what the Tevatron gives (see e.g.~\cite{ua2pheno,Cheung:2011zt}). 
For a $\zp $ boson with $\sim 144$ GeV mass, the UA2 bound on the 
$\zp $ coupling to quarks is estimated in 
Ref.~1~of~\cite{ua2pheno}.
Their results are consistent with our analysis given below. \\

As discussed after Eq.(\ref{r31}), we assume that the $\zp$ couples mostly to quarks and we will assume a $\zp$ mass of $144\GeV$ and a coupling of $g_X=0.35$.
 For simulations we use {\tt MadGraph~4.4}~\cite{madgraph}, {\tt PYTHIA}~\cite{pythia} and {\tt PGS 4}~\cite{pgs} and we consider the $Wjj$, $Zjj$ and $\gamma jj$ production channels  where the dijets arise from the Standard Model (SM) or from the decay of the $\zp$.    All processes are simulated at $\sqrt{s}=1.96~{\rm TeV}$ for a $p\bar{p}$ collider,  at $\sqrt{s}=7~{\rm TeV}$ for a $pp$ collider (LHC7) and the $\zp \to jj$ production is simulated at UA2 ($p\bar{p}$ collider with $\sqrt{s}=630\GeV$) to verify that this model was not already excluded~\cite{ua2}.  At the Tevatron and LHC7, we do not consider the single production of $\zp$, i.e. $p\bar{p}\to\zp\to jj$ or $pp\to\zp\to jj$, since this would be a relatively hard signal to find compared to the standard model, namely due to QCD.  The contribution of this $\zp$ model to the $Wjj$ (pre-cut) cross section is $3.62\pb$, which is in agreement with the CDF reported value~\cite{CDF} and the number of events (after cuts) are in good agreement with the CDF reported values~\cite{CDF} as shown in Table~\ref{numevents}. The effective cross section of $Wjj$, $Z jj$ and $\gamma jj$ after trigger and cut efficiencies are taken into account at the Tevatron and LHC7 are shown in Table~\ref{tab1} and Table~\ref{tab2}.  
Below we discuss our selection cuts on various final states in details.\\

In Ref.~1~of~\cite{ua2}, the UA2 Collaboration puts a $90\%$~CL upper limit on the dijet production rate of an extra vector boson.  For our $\zp$ model we calculate the cross section (taking into account the cut and the trigger efficiencies reported in Ref.~1~of~\cite{ua2}) 
to be $2.32\times10^{2}\pb$.  As pointed out in~\cite{nelson}, the analysis of UA2 was done using comparatively primitive Monte Carlo, detector simulations and jet algorithms.  For these reasons we assume, as in~\cite{nelson}, that the UA2 bound is an order of magnitude limit. \\

The particle identification criteria used for the Tevatron are as follows: a lepton~(electron or muon) candidate must have $p_{T}>20\GeV$ and $\left|\eta\right|<1.0$. Jets candidates have $p_{T}>30\GeV$, $\left|\eta\right|<2.4$ and are removed if the jet is within $\Delta R=\sqrt{\left(\Delta \eta\right)^2+\left(\Delta\phi\right)^2}<0.52$ of a lepton.  Following the framework of~\cite{Cheung:2011zt}, for the $\gamma jj$ search we use the criteria that the selected photon must have $p_{T}>50\GeV$ and $\left|\eta\right|<1.1$, which is a higher momentum cut than the one used in~\cite{Wjjzaref}.  Similar identification criteria are used for the LHC7 analysis.\\

For the $Wjj$ analysis we follow the cuts used in~\cite{CDF} where events are selected to have one identified lepton, two identified jets and missing transverse energy.  In addition to this, we check to make sure the event does not have a second lepton (with
 $p_{T}>10\GeV$) and that the dilepton invariant mass is not  in the $Z$ range.  Further selection includes events with two
 identified jets, missing transverse energy which exceeds $25\GeV$ and that the transverse mass between the lepton and missing transverse energy exceeds $30~\GeV$. The two jets must
  have $\left|\Delta\eta\right|<2.4$ and  the momentum of the dijet system must exceed $40\GeV$.  In addition events are required to have the spacing between the missing transverse
   energy and the leading jet to be separated by at least $\left|\Delta\phi\right|=0.4$.  After applying these cuts we calculate that our model produces $104\pm10$ electron events and
    $124\pm11$ muon events which are roughly within $1\sigma$ of the values CDF reported~\cite{CDF}.  Table~\ref{numevents} shows how our model compares to the CDF values.   \\
    
\begin{table}
\begin{center}
{\bf Number of Signal Events for  $Wjj$ at the Tevatron}\\
\vspace{0.2 cm}
\begin{tabular}{l||c|c}
& CDF~(events)& $\zp$~(events) \tnhl
Electron & $156\pm42$ & $104\pm10$\\
Muon & $97\pm38$ & $124\pm11$ \\
\end{tabular}
\caption{\label{numevents} Exhibition of the number of excess events in the electron and muon channels  at the Tevatron for $Wjj$ for the model discussed in the text as well as the CDF reported value~\cite{CDF}.  The displayed values are after cuts and the uncertainty shown 
 for the $\zp$ model is only statistical and does not take into account systematic uncertainties. }
\end{center}
\end{table}

Now for the $Zjj$ analysis, we select events with two identified leptons with the dilepton invariant mass  in the range of $76\GeV$ to $106\GeV$, i.e. the $Z$~range.  Further events are required to have two identified jets with the same jet event selection as the $Wjj$ case including the dijet system momentum and $\Delta\eta$. After taking into account the trigger and cut efficiencies, the cross section of the signal is $3.8~{\rm fb}$ compared to 
$213.5~{\rm fb}$ for the SM background, which at the Tevatron with $4.3~{\rm fb}^{-1}$ of integrated luminosity would not produce a visible excess. Additionally, if one also requires the dijet invariant mass to be in the $120\GeV$ to $160\GeV$ range the  effective cross section of the signal becomes $0.8~{\rm fb}$.\\

Additionally, events are selected for the $\gamma jj$ analysis that have no identified leptons, one identified photon, 
dijet invariant mass in the range of $120\GeV$ to $160\GeV$ and two identified jets with the same jet event selection as the $Wjj$ case.  The cross section for the signal after trigger and cut efficiencies is $72.1~{\rm fb}$ and for the SM background we get  $3.0\times10^{3}~{\rm fb}$, which at the Tevatron with $4.3~{\rm fb}^{-1}$ of integrated luminosity would not produce a visible excess.\\

\begin{table}[h!]
\begin{center}
{\bf   Effective Dijet Cross Sections at the Tevatron}\\
\vspace{0.2 cm}
\begin{tabular}{l||c|c}
& SM~$({\rm fb})$ & $\zp$~$({\rm fb})$ \tnhl
$Wjj$ & $3.2\times 10^{3}$ & $53.1$\\
$Zjj$ & $213.5$ & $3.8$ \\
$\gamma jj$& $3.0\times 10^{3}$ & $72.1$\\
\end{tabular}
\caption{\label{tab1} Exhibition of the effective cross sections at the Tevatron for $Wjj$, $Zjj$ and $\gamma jj$ using the $\zp$ model (as well as the cuts) discussed in the text.  Before cuts the $\zp$ contribution to the $Wjj$ cross section is $3.62\pb$ and is in agreement with the CDF reported value~\cite{CDF}.  As shown in Table~\ref{numevents}, the number of events for this $\zp$ model agrees within $1\sigma$ to the CDF reported value~\cite{CDF}.
}
\end{center}
\end{table}

We discuss now the implications of the model at LHC7 using the same trigger and cut efficiencies as stated above.
 For the $Wjj$ production channel we find that the signal cross section is $160.5~{\rm fb}$ compared to the SM cross section of $3.4\times 10^{4}~{\rm fb}$.
After applying the $Zjj$ analysis we find that the  effective cross section for the signal  is $9.3~{\rm fb}$ and the SM background is $2.4\times10^{3}~{\rm fb}$.  If we further require that the dijet invariant mass be in the $120\GeV$ to $160\GeV$ range we get the effective cross section of the signal to be $1.5~{\rm fb}$.
  Additionally, for the $\gamma jj$ channel we find the  effective cross section to be $115.5~{\rm fb}$ for the signal and $6.2\times 10^{3}~{\rm fb}$ for the SM background which gives  a $5\sigma$  excess with $12.3~{\rm fb}^{-1}$ of integrated luminosity.  For this part of the analysis we have used $S/\sqrt{B}=5$, where $S$ is the number of signal events and $B$ is the number of background events; however with a better statistical procedure and/or better set of cuts a possible discovery could occur at a lower luminosity~\cite{atlasTDR}.\\

\begin{table}[h!]
\begin{center}
{\bf Effective Dijet Cross Sections at LHC7}\\
\vspace{0.2 cm}
\begin{tabular}{l||c|c}
& SM~$({\rm fb})$ & $\zp$~$({\rm fb})$\tnhl
$Wjj$ & $3.4\times 10^{4}$ & $160.5$\\
$Zjj$  & $2.4\times 10^3$& $9.3$ \\
$\gamma jj$& $6.2\times 10^{3}$ & $115.5$\\
\end{tabular}
\caption{\label{tab2} Display of the effective cross sections at the LHC with $\sqrt{s}=7~{\rm TeV}$ for $Wjj$, $Zjj$ and $\gamma jj$ using the $\zp$ model and cuts discussed in the text.}
\end{center}
\end{table}

In the above analysis we have ignored the corrections arising from finite but small $\epsilon=M_2/M_1$.
Inclusion of this term would make only a small correction relative to the 
contribution arising from $J_X^{\mu}$ in the hadronic channels
 and thus all our conclusions above remain unchanged.
 The decay width of such a $\zp$ into quarks is given by
 \begin{equation}
 \Gamma\left(\zp\to q\bar{q}\right)=\frac{N_cN_fg_X^2}{12\pi}M_{\zp}\left(1+\frac{\alpha_{s}}{\pi}\right)~,
 \end{equation}
 where $N_c=3$ is the number of colors and $N_f$ is the number of flavors ($N_{f}=5$ for $M_{\zp}=144\GeV$) which gives $\Gamma\left(\zp\to q\bar{q}\right)\simeq7.3\GeV$.  If we turn on the mixings between $B_{\mu}$ and $X_{\mu}$ then such mixing is constrained by the precision electroweak data.  We have analyzed the constraints on $\epsilon$ from the electroweak data and find that these constraints are much less stringent than in the analysis of~\cite{kn,fln} since here $X_{\mu}$ couples only to quarks.  Our analysis shows that for the present model $\epsilon<0.11$ compared to the more stringent constraint of $\epsilon<0.05$ in the works of~\cite{kn,fln}.  
 Now  a  small $M_2/M_1$ would produce  a  small production cross section for leptons 
  via the Drell-Yan process $p\bar p\to \zp \to \ell^+\ell^-$. A Stueckelberg $\zp $ in the dileptonic 
channel has been probed by the D\O\  experiment at the Tevatron~\cite{Abazov:2010ti} which put
limits on $\epsilon$ for various $\zp $ masses. Thus the experiment puts a constraint
on $\epsilon$ so that $\epsilon <0.02$ for $M_{\zp } =200$ GeV.  We estimate that for $\zp $ mass
of 150 GeV, the limit on $\epsilon$  from the D\O\ experiments would be smaller than $0.02$.  Thus the Tevatron gives a more stringent limit on $\epsilon$ than the precision electroweak analysis. \\

In conclusion, we have analyzed in this work the $Wjj$ anomaly reported by the CDF experiment at the Tevatron. 
We show that the dijet anomaly can arise from  a  $U(1)_X$ Stueckelberg extension of the Standard Model
where the $U(1)_X$ couples only with the quarks. An extra $U(1)$ gauge field coupling to quarks only 
was also considered in \cite{Cheung:2011zt,perezetal}.    However, unlike the analysis of~\cite{Cheung:2011zt}
our couplings are purely vector and we work within the Stueckelberg mechanism where no Higgs
is required for the $\zp $ mass growth. Our framework is also different from of~\cite{perezetal}.
We have also analyzed the $Zjj$ and $\gamma jj$ production and find that they are consistent 
with current data as well as the results reported in~\cite{Cheung:2011zt,Wjjzaref}.  We further extend our results to show expected effective cross sections at the LHC assuming the same event selection and identification criteria as at the Tevatron. Thus the $U(1)_X$ Stueckelberg extension of the Standard Model appears
a valid  explanation of the $Wjj$ anomaly if such an anomaly is indeed confirmed by D{\O},  the LHC or by further data. 
Finally we discuss some distinguishing features of the Stueckeberg model proposed here from other  $Z'$ models.
The Stueckelberg couplings are purely vector like and in this sense the model is very different from other
$Z'$ models where in general there is a combination of vector and axial vector couplings. While  the 
current data on the Wjj anomaly is unable to discriminate between the Stueckelberg and other
models, it would be possible to discriminate among models from forward-backward asymmetry if such
asymmetry can be measured in future data. Further, the model produces small  branching ratio
of $Z'$ into two leptons. Again while this is not discernible in current data such small effects 
could be used to discriminate the Stueckelberg from other baryonic  $U(1)'$  models in future data if such data
becomes available. Finally there is no residual
Higgs field in this case which would be the case if the $Z'$ mass was generated by the normal
Higgs phenomena. 
\\


\noindent {\it Acknowledgements:} 
Discussions with Hooman Davoudiasl, Daniel Feldman, Patrick Meade, Amarjit Soni, and Darien Wood are acknowledged.
This research is  supported in part by the U.S. National Science Foundation (NSF) grants
 PHY-0757959 and PHY-0969739 and through the TeraGrid under grant numbers TG-PHY110015 and TG-PHY100036.

\end{document}